\documentclass[twocolumn,prb,superscriptaddress,psfig,showpacs,amsmath,amssymb,fleqn]{revtex4-1}
\setcitestyle{square,numbers}

\usepackage{graphicx}
\usepackage[caption=false]{subfig}
\usepackage{epstopdf}
\usepackage{dcolumn}
\usepackage{bm}
\usepackage{hyperref}
\hypersetup{
	colorlinks   = true, 
	urlcolor     = blue, 
	linkcolor    = blue, 
	citecolor    = blue 
}
\usepackage{xspace}
\usepackage{color}
\usepackage{soul}

\newcommand{\NCCF}{NaCaCo$_2$F$_7$\xspace}

\newcommand{\etal}{\textit{et al.}\xspace}

\newcommand{\kB}{\ensuremath{k_\mathrm{B}}\xspace}

\newcommand{\muB}{\ensuremath{\mu_\text{B}}\xspace}
\newcommand{\mueff}{\ensuremath{\mu_{\mathrm{eff}}}\xspace}

\newcommand{\ETO}{Er$_2$Ti$_2$O$_7$\xspace}
\newcommand{\eee}{$\langle$111$\rangle$\xspace}

\begin{document}

\title{Magnetic interactions and spin dynamics in the bond-disordered pyrochlore fluoride NaCaCo$_2$F$_7$}
\author{J.~Zeisner}
\altaffiliation[]{j.zeisner@ifw-dresden.de}
\affiliation{Leibniz Institute for Solid State and Materials Research IFW Dresden, D-01069 Dresden, Germany}
\affiliation{Institute for Solid State and Materials Physics, TU Dresden, D-01069 Dresden, Germany}
\author{S.~A.~Br\"auninger}
\altaffiliation[]{sascha\_albert.braeuninger@tu-dresden.de}
\affiliation{Institute for Solid State and Materials Physics, TU Dresden, D-01069 Dresden, Germany}
\author{L.~Opherden}
\altaffiliation[]{l.opherden@hzdr.de}
\affiliation{Dresden High Magnetic Field Laboratory (HLD-EMFL), Helmholtz-Zentrum Dresden-Rossendorf, D-01328 Dresden, Germany}
\author{R.~Sarkar}
\affiliation{Institute for Solid State and Materials Physics, TU Dresden, D-01069 Dresden, Germany}
\author{D.~I.~Gorbunov}
\affiliation{Dresden High Magnetic Field Laboratory (HLD-EMFL), Helmholtz-Zentrum Dresden-Rossendorf, D-01328 Dresden, Germany}
\author{J.~W.~Krizan}
\affiliation{Department of Chemistry, Princeton University, Princeton, NJ 08544, USA}
\author{T.~Herrmannsd\"{o}rfer}
\affiliation{Dresden High Magnetic Field Laboratory (HLD-EMFL), Helmholtz-Zentrum Dresden-Rossendorf, D-01328 Dresden, Germany}
\author{R.~J.~Cava}
\affiliation{Department of Chemistry, Princeton University, Princeton, NJ 08544, USA}
\author{J.~Wosnitza}
\affiliation{Institute for Solid State and Materials Physics, TU Dresden, D-01069 Dresden, Germany}
\affiliation{Dresden High Magnetic Field Laboratory (HLD-EMFL), Helmholtz-Zentrum Dresden-Rossendorf, D-01328 Dresden, Germany}
\author{B.~B\"{u}chner}
\affiliation{Leibniz Institute for Solid State and Materials Research IFW Dresden, D-01069 Dresden, Germany}
\affiliation{Institute for Solid State and Materials Physics, TU Dresden, D-01069 Dresden, Germany}
\author{H.-H.~Klauss}
\affiliation{Institute for Solid State and Materials Physics, TU Dresden, D-01069 Dresden, Germany}
\author{V.~Kataev}
\affiliation{Leibniz Institute for Solid State and Materials Research IFW Dresden, D-01069 Dresden, Germany}
\date{\today}

\begin{abstract} 
We report high-frequency/high-field electron spin resonance (ESR) and high-field magnetization studies on single crystals of the bond-disordered pyrochlore \NCCF. Frequency- and temperature-dependent ESR investigations above the freezing temperature \mbox{$T_f \sim 2.4$\,K} reveal the coexistence of two distinct magnetic phases. A cooperative paramagnetic phase, evidenced by a gapless excitation mode, is found as well as a spin-glass phase developing below 20\,K which is associated with a gapped low-energy excitation. Effective $g$-factors close to 2 are obtained for both modes in line with pulsed high-field magnetization measurements which show an unsaturated isotropic behavior up to 58\,T at 2\,K. In order to describe the field-dependent magnetization in high magnetic fields, we propose an empirical model accounting for highly anisotropic ionic $g$-tensors expected for this material and taking into account the strongly competing interactions between the spins which lead to a frustrated ground state. As a detailed quantitative relation between effective $g$-factors as determined from ESR and the local $g$-tensors obtained by neutron scattering \mbox{[Ross \textit{et al.}, Phys. Rev. B \textbf{93}, 014433 (2016)]} is still sought after, our work motivates further theoretical investigations of the low-energy excitations in bond-disordered pyrochlores.
\end{abstract}

\maketitle

\section{\label{sec:introduction} Introduction}
Strongly frustrated spin systems incessantly attract considerable attention within the solid-state research community as they show a variety of peculiar electronic properties. Among them are, for example, unusual ground states such as quantum-spin-liquid, spin-ice, and spin-glass states as well as exotic excitations such as magnetic monopoles in spin-ice compounds \cite{Balents2010}. In many cases, frustration arises from the combination of a specific lattice structure, e.g., a triangular or a kagome lattice, with strong (antiferromagnetic) exchange interactions, which is referred to as geometrical frustration \cite{Ramirez1994}. Pyrochlore oxides with the general formula $A_2B_2$O$_7$ belong to the prime examples for realization of this type of magnetic frustration because their $A$ and $B$ site sublattices consist of corner-sharing tetrahedra. In addition to fulfilling the prerequisite for geometrical frustration, pyrochlore oxides could be successfully synthesized with a plethora of combinations regarding the magnetic, in particular, rare-earth ions which led to a boost of experimental and theoretical research activities on these materials during the last decades \cite{Gardner2010}. Due to the possibility to interchange magnetic ions, pyrochlores offer a vast materials base for exploring the properties of spin systems with different types of exchange interactions, for instance, Heisenberg, $XY$, or Ising-type interactions on a frustrated lattice. Similar to the other pyrochlores, the group of $XY$ pyrochlores shows a broad phenomenology regarding their ground states due to a competition of several phases at low temperatures \cite{Hallas2018}. For example, a long-range antiferromagnetically ordered state was found in \ETO potentially driven by the order-by-disorder mechanism whereas a ferromagnetically ordered state with high sensitivity to small amounts of disorder was observed in Yb$_2$Ti$_2$O$_7$, see, e.g., Refs.~\cite{Hallas2018,Yan2017,Bhattacharjee2016} and references therein. 

The system which is studied here, \NCCF, is proposed to belong to the class of $XY$ pyrochlores \cite{Ross2016}. It is one of the first members of the transition-metal fluorine-based pyrochlore family \cite{Haensler1970} which came into focus of interest in recent years due to the availability of large single crystals \cite{Krizan2014,Krizan2015,Krizan2015a,Sanders2017}. In contrast to the majority of pyrochlore oxides, magnetic species in the fluoride compounds are $3d$ transition metal ions which leads to stronger interactions between the spins \cite{Hallas2018}. Furthermore, while the $B$ sites are uniformly occupied by Co$^{2+}$ ions centered in compressed fluorine octahedra, the $A$ site sublattice hosts Na$^+$ and Ca$^{2+}$ ions which are randomly distributed over the respective sites \cite{Krizan2014} (see, e.g., Fig. 1 in Ref.~\cite{Krizan2014} and Fig. 1 in Ref.~\cite{Ross2017}). The induced disorder at the non-magnetic sublattice influences the crystal electric fields (CEF) at the magnetic sites as well as the exchange paths between the Co ions, thereby leading to disorder in the (effective) exchange interactions. Theoretical studies of the impact of such quenched disorder on the ground state of $XY$ pyrochlores revealed that, depending on the degree of disorder, either a long-range ordered state is selected from the manifold of competing phases or a spin-glass phase is established above a critical level of disorder \cite{Andrade2018}. Indeed, initial characterizations of \NCCF showed a spin freezing at a temperature $T_f \sim 2.4$\,K, although a Curie-Weiss analysis of the magnetic susceptibility yields strong antiferromagnetic (AFM) interactions evidenced by a Curie-Weiss temperature \mbox{$\Theta_{CW} =  -140$\,K} \cite{Krizan2014}. Furthermore, recent inelastic neutron scattering (INS) studies found a short-range ordered state below $T_f$ consisting of $XY$-like AFM clusters with the associated spin fluctuations persisting even to temperatures above $T_f$ \cite{Ross2016}. In addition, based on INS experiments a single-ion model was proposed for \NCCF featuring an effective $J_{\rm eff} = 1/2$ description of the magnetic ions with a highly anisotropic $XY$-type $g$-tensor which contains, in addition, a small Ising-contribution \cite{Ross2017}. However, measurements of static magnetization did not show an anisotropic behavior \cite{Ross2016}, as might be anticipated from the reported $g$-tensors, pointing towards more complex low-temperature properties of this bond-disordered pyrochlore. 

Here, we address the question of the apparently isotropic behavior and investigate, in addition, the low-energy excitations above $T_f$. High-field/high-frequency electron spin resonance (HF-ESR) experiments at temperatures between 3 and 40\,K yield evidence for the coexistence of a cooperative paramagnetic and a spin-glass phase. ESR modes related to these two phases have effective \mbox{$g$-factors} around 2 which is in line with the isotropic high-field magnetization measurements in fields up to 58\,T at 2\,K. This paper is organized as follows. In Sec.~\ref{sec:exp_details} details on the samples employed in these studies and experimental methods are provided. The results of our experiments are presented in Sec.~\ref{sec:exp_results} followed by their discussion in Sec.~\ref{sec:Discussion}, where among others an empirical model for the magnetization data is proposed. The main conclusions of this work are summarized in Sec.~\ref{sec:conclusions}. Additional information on the samples used in the magnetization experiments can be found in Appendix~\ref{sec:sample_details}. 
 
\section{\label{sec:exp_details} Samples and Experimental Methods}
Single crystals of \NCCF were grown using the floating-zone technique as described in  detail in Ref.~\cite{Krizan2014}. In order to reduce the influence of demagnetization effects on the magnetization measurements, two cube-shaped samples were cut out of a large single crystal for these experiments, see Appendix~\ref{sec:sample_details}.

HF-ESR studies were performed employing a homemade setup. The sample was placed into a magneto-cryostat (Oxford Instruments) which enabled measurements in magnetic fields up to 16\,T and at temperatures between 2~and 300\,K. Microwaves with frequencies up to 480\,GHz were generated using an amplifier/multiplier chain (AMC from Virgina Diodes Inc.) and were detected by a hot-electron InSb bolometer (QMC Instruments Ltd.). In addition, a vector network analyzer (PNA-X from Keysight Technologies) was utilized for generation and detection of microwaves in a frequency range from 75~to~330\,GHz. All ESR spectra were recorded in transmission using a Faraday configuration and applying the external magnetic field along the crystallographic [111] direction. Crystallographic directions were determined by single crystal x-ray diffraction \mbox{(see Appendix~\ref{sec:sample_details}).}

Field-dependent magnetization measurements in pulsed magnetic fields up to 58\,T and at temperatures between 2 and 129\,K were conducted at the Dresden High Magnetic Field Laboratory using an induction method with a pickup-coil device. The absolute values of the magnetization were calibrated using static-field data. A detailed description of the high-field magnetometer is given in Ref.~\cite{Skourski2011}.
\section{\label{sec:exp_results} Results}

\subsection{\label{sec:results_ESR} Electron spin resonance}
\begin{figure}[t!]
	\includegraphics[width=\columnwidth]{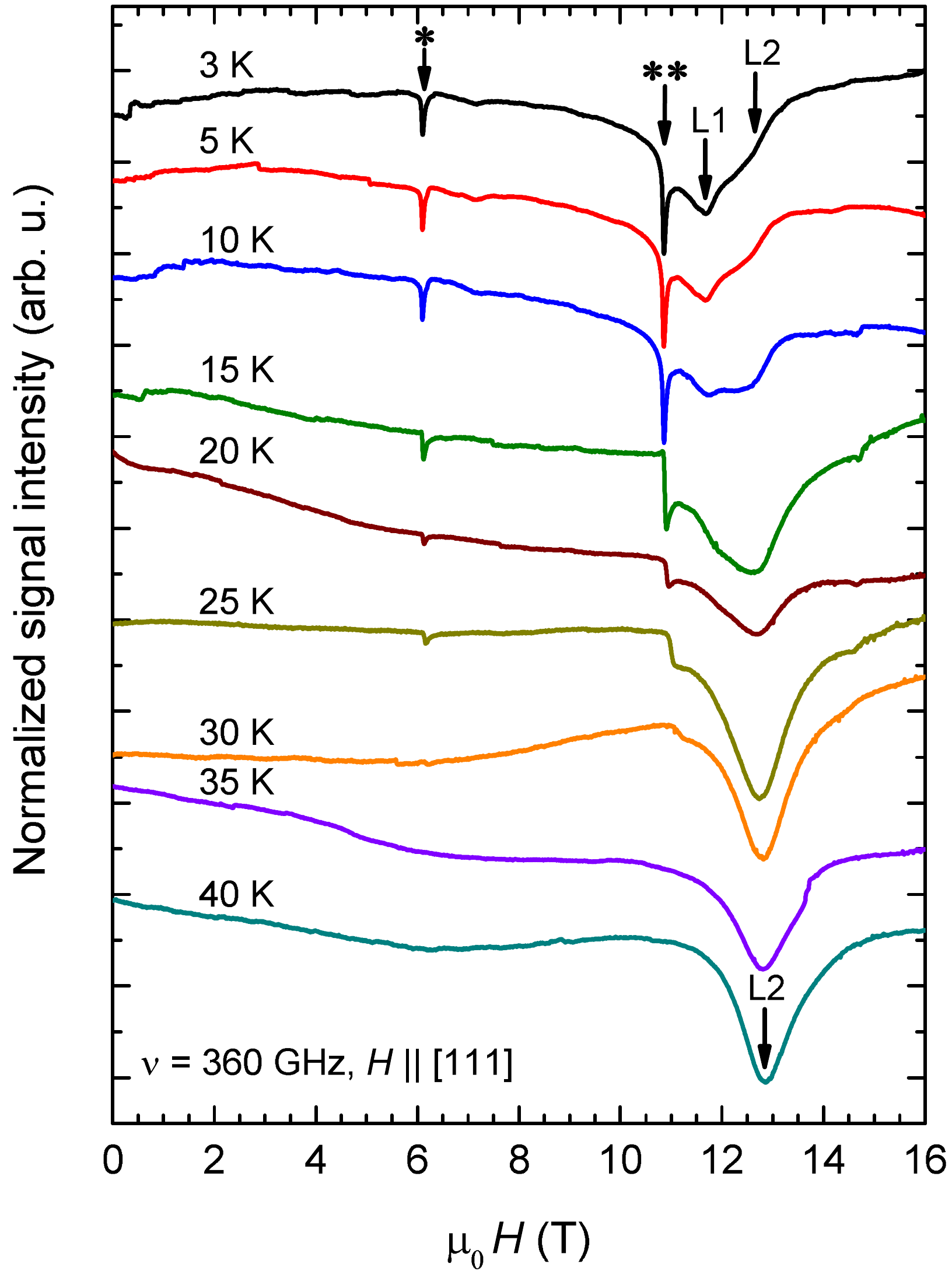}
	\caption{Temperature evolution of the ESR spectrum at 360~GHz for an external magnetic field applied along the crystallographic [111] direction. Shown spectra are normalized and shifted vertically for better comparison.}
	\label{fig:ESR_spectra_temp}
\end{figure}

In the following, results of our ESR investigations are presented, starting with the temperature dependence of the signal. Measurements were carried out for temperatures between 2 and 200\,K. However, resonance lines could be observed between 3~and~40\,K only. In Fig.~\ref{fig:ESR_spectra_temp}, spectra recorded within this temperature range are shown for a fixed frequency $\nu$ of 360\,GHz. At 40\,K, the spectrum consists of a single resonance line, denoted as L2 in Fig.~\ref{fig:ESR_spectra_temp}, which is centered at about 12.8\,T and has a linewidth (full width at half maximum) of approximately 1.3\,T. Upon cooling, the spectrum becomes more structured: starting at 30\,K two additional peaks, marked with asterisks ($\ast$) and ($\ast \ast$) in Fig.~\ref{fig:ESR_spectra_temp}, appear in the recorded spectra. These peaks are artifacts originating from molecular oxygen contained in small amounts of solidified air in the waveguide, see, for comparison, Ref.~\cite{Pardi2000}, and are not related to the ESR response of \NCCF. Below 20\,K, an additional line, L1, appears in the vicinity of L2. Further lowering the temperature results in a change of the relative heights of the peaks. We note that, due to experimental reasons, no absolute values of intensities are measured which only allows conclusions concerning changes of the peak heights relative to each other. Compared to L2, line L1 grows in height until L2 is visible only as a shoulder of L1 at 3\,K. Within the whole temperature interval discussed here, the observed peaks do not show a pronounced temperature dependence of their positions. Thus, it may be concluded that L1 and L2 at low temperatures are indeed two separate peaks, with presumably different origins, and are not the result of a splitting of the line L2 as seen at elevated temperatures, i.e., above 20\,K.
 
The absence of an ESR signal below $T_f$ could be related to the high degree of static disorder in the frozen state which extends over the whole crystal below the freezing transition. Strong disorder can give rise to a broad distribution of internal fields and, thus, to a large inhomogeneous broadening of resonance lines, as it was observed for other Co-based disordered spin systems, see, for instance, Refs. \cite{Iakovleva2015,Iakovleva2017}. Indeed, such a broad distribution of static internal fields below $\sim 3.6$\, K was evidenced by the broadening of the NMR lines in Ref.~\cite{Sarkar2017}. The combination of a strongly broadened line and a low overall signal intensity, determining the area under the absorption curve, leads to a rather flat ESR line which is undetectable with the employed setup.

\begin{figure}
	\centering
	\includegraphics[width=\columnwidth]{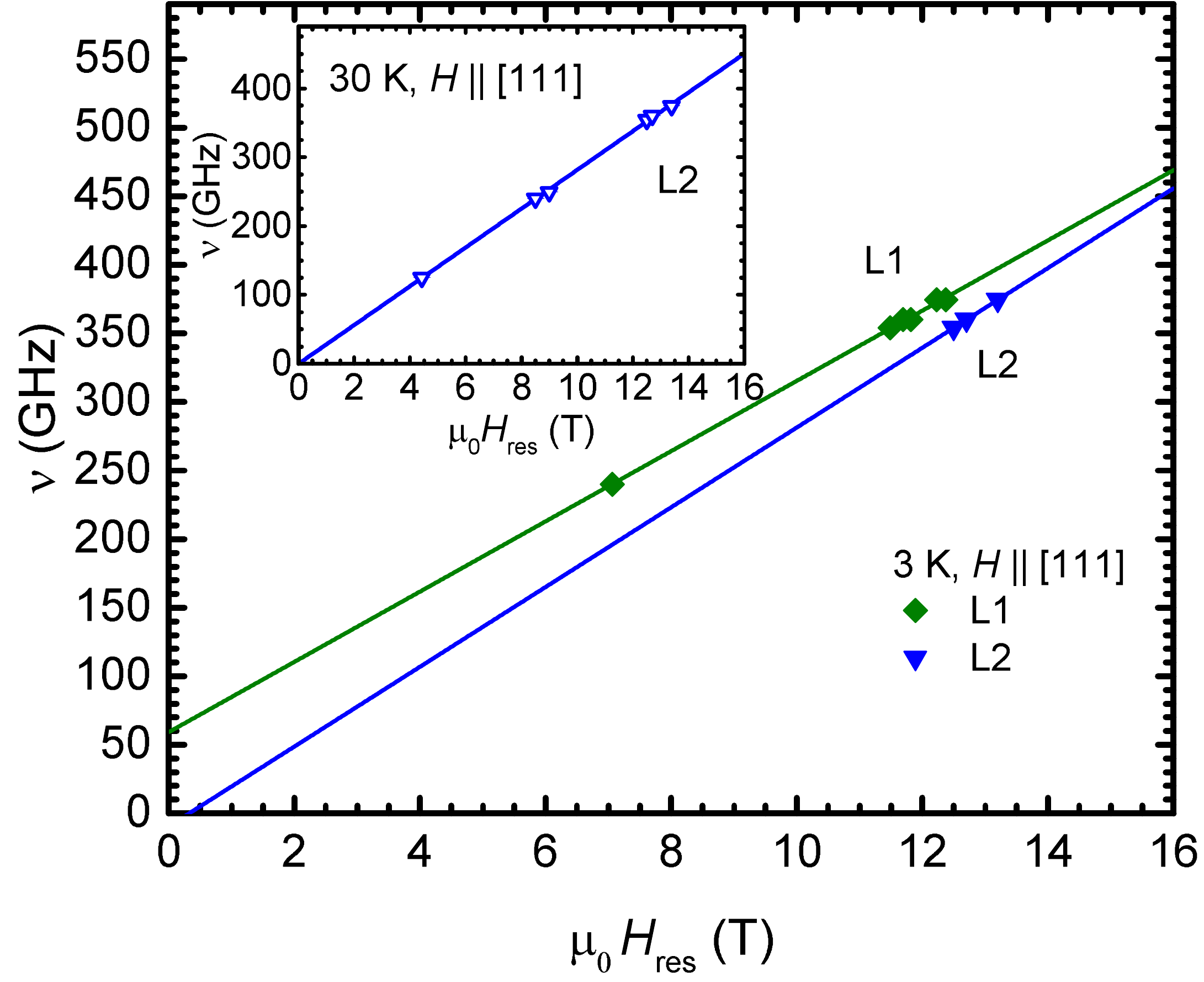}
	\caption{Frequency-field diagrams obtained from measurements at 3~K (main panel) and 30~K (inset) with external magnetic field applied along the crystallographic [111] direction. Solid lines are fits to Eq. (\ref{eq:linear}) as described in the main text.}
	\label{fig:freq-dep}
\end{figure}

In order to investigate the nature of the various observed resonance lines, frequency-dependent measurements were conducted at 3 and 30\,K. The corresponding frequency-field diagrams are shown in Fig.~\ref{fig:freq-dep}. On the phenomenological level, for both lines the frequency-field dependence could be described by the following expression:
\begin{equation}
\nu = g\mu_{\text{B}}\mu_{\text{0}} H_{\text{res}} / h + \Delta \ \ .
\label{eq:linear}
\end{equation} 
Here $g$ denotes the (effective) $g$-factor of the magnetic entities responsible for the observed lines, $H_{\text{res}}$ is the resonance field of the peaks, and $\Delta$ describes the anisotropy/excitation gap. The constants $\mu_{\text{B}}, \ \mu_{\text{0}}$, and $h$ are Bohr's magneton, the vacuum permeability, and Planck's constant, respectively. Equation (\ref{eq:linear}) describes the frequency-field dependence for the case of antiferromagnetic resonance (AFMR) if $\Delta \neq 0$ \cite{Turov}, while for $\Delta = 0$ the usual expression for paramagnetic resonance is recovered. Note that Eq.~(\ref{eq:linear}) describes the resonance condition of an ordered collinear two-sublattice antiferromagnet for the case of an external magnetic field applied along the ``easy'' direction \cite{Turov}. Although the title compound does not show true long-range antiferromagnetic order at low temperatures \cite{Ross2016}, these relations nonetheless provide a satisfactory description of the experimentally observed frequency-field dependences. Fit results for $g$ and $\Delta$ are summarized in Table~\ref{tab:fit_results}. Line L2, which dominates the spectra at 30\,K, has at both temperatures a $g$-factor of about 2, with the $g$-factor being slightly larger at 3\,K. Moreover, analysis of the frequency-field dependence evidences a gapless behavior at 30\,K, whereas the situation is less clear at 3\,K due to the limited amount of data points caused by the merging of lines L1 and L2, and the fact, that these lines are observed at high frequencies only. Thus, the gap value obtained from fit of Eq.~(\ref{eq:linear}) to the data of L2 at 3\,K is not reliable and is, most likely, zero. We note that forcing the fit to go through the origin would yield a $g$-factor of 2.03(1). Therefore, the small change in $g$-factor might also be attributed to this uncertainty and is not considered in the following. In contrast to L2, the frequency-field dependence of L1 reveals a non-zero gap value. At 3\,K, the behavior of L1 follows Eq. (\ref{eq:linear}) with a $g$-factor being smaller than two, see Table \ref{tab:fit_results}.

\begin{table}[b!]
	\caption{Results for $g$ and $\Delta$ obtained from frequency-dependent measurements at various temperatures by fitting Eq. (\ref{eq:linear}) to the data. Values given in parentheses are uncertainties of the respective fits.}
	\setlength\extrarowheight{0.5pt}
	\begin{ruledtabular}
		\begin{tabular}{ c  c  c  c  c }
			&  $T$ = 3\,K & &  $T$ = 30\,K &  \\
			Line & $g$ & $\Delta$ (GHz) &$g$ & $\Delta$ (GHz) \\
			\colrule 
			L1 & $1.83(3)$ & $59(4)$& - & -  \\
			L2 & $2.08(2)$ & $-10(10)$ \footnotemark &  $2.01(1)$& $0$ \\
		\end{tabular}
	\end{ruledtabular}
	\footnotetext[1]{This value given as uncertainty of $\Delta$ is larger than the uncertainty of the fit (4\,GHz) due to the limited amount of data points available, see text.}
	\label{tab:fit_results}
\end{table}

\subsection{\label{sec:results_Magnetization} Magnetization}

\begin{figure*}[!htbp]
	\centering
	\includegraphics[width=\textwidth]{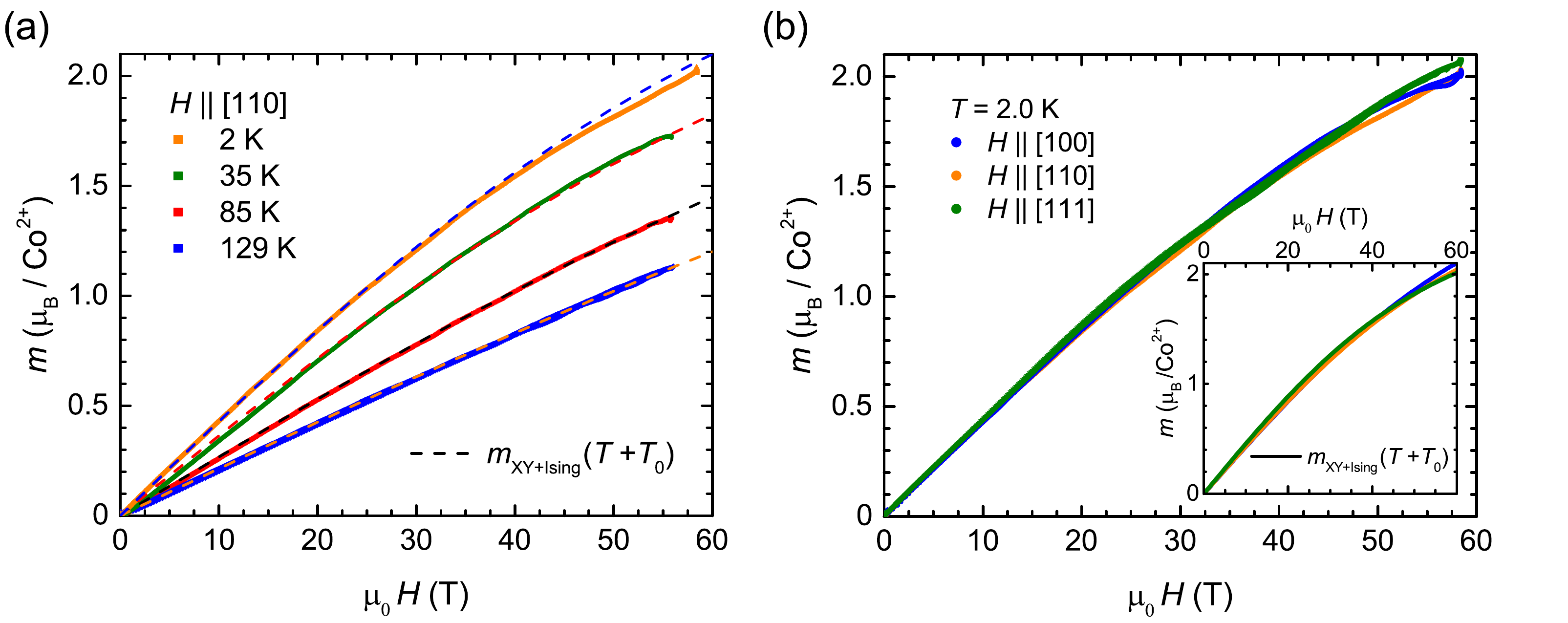}
	\caption{Field dependence of the magnetization of \NCCF at various temperatures with $H || [110]$ (a) and for different orientations at 2\,K (b). Dashed lines in (a) and solid lines in the inset of (b) are fits to the data assuming an effective temperature $T_\mathrm{eff} = T + T_0$ as well as a small Ising contribution ($\mu_\mathrm{Ising}$ = 0.935 \muB) in addition to the dominant $XY$ moments ($\mu_\mathrm{xy}$ = 3.040 \muB), see text for details.}
	\label{fig:NCCF_02}
\end{figure*}

Measurements of the magnetization in low static \cite{Krizan2014,Ross2016} as well as high pulsed fields (Fig. \ref{fig:NCCF_02}) underline the existence of strong antiferromagnetic interactions which lead to the strong magnetic frustration in the pyrochlore system. Field dependence of the magnetization was measured at various temperatures between 2 and 129\,K with the external magnetic field applied parallel to the $[110]$ direction, see Fig.~\ref{fig:NCCF_02}(a). Even at a temperature of 2\,K and a magnetic field of 58\,T full polarization of the spin system is not reached. In comparison, for non-interacting Heisenberg spins almost full saturation is expected for a magnetic field of a few Tesla.

Furthermore, the field dependence of the magnetization was studied at 2\,K for three different orientations of the external magnetic field, Fig.~\ref{fig:NCCF_02}(b), in order to shed light on the magnetic anisotropies present in the title compound. A local $XY$ anisotropy for spins on a lattice of corner-sharing tetrahedra, as might be inferred from the crystal structure \cite{Krizan2014}, should naively lead to a different behavior for different field directions. The magnetization of \NCCF, however, shows no strong anisotropy which is consistent with magnetization measurements at low fields reported previously \cite{Ross2016}. This is in contrast to the known pyrochlore system with $XY$ anisotropy, \ETO. In this case a feature is seen in the field-dependent magnetization at about 1\,T which shows the expected difference in the absolute value for the three main directions of the cubic system \cite{Petrenko2011,Petrenko2013, Lhotel2017}.
The absence of such a difference in \NCCF might be caused by an accidental \textquotedblleft masking\textquotedblright \ of $XY$ anisotropy. As discussed in more detail in Sec.~\ref{sec:discussion_Magnetization}, the occurrence of a small Ising contribution and, in particular, of strong competing spin-spin interactions affects the spin configuration with simple $XY$ anisotropy.

\section{\label{sec:Discussion} Discussion} 

In the preceding section we presented the results of a combined ESR and magnetization study aiming at a detailed investigation  of the magnetic properties and spin dynamics of the frustrated pyrochlore system \NCCF. First, we discuss our findings by proposing a possible scenario for the origin of the different resonance lines found in ESR experiments. Then, we turn to a discussion of the isotropic behavior observed in the high-field magnetization measurements using an empirical model to consider interactions between the spins.

\subsection{Electron spin resonance}
\label{sec:discussion_ESR}
As was mentioned in the introduction of this paper, the magnetic ions in this system are Co$^{2+}$ ions placed in a distorted octahedral environment (see, e.g., Fig.~1 in Ref.~\cite{Krizan2014}). Such an arrangement typically leads to highly anisotropic $g$-tensors and a ground state characterized by an effective total angular momentum $J_{\rm eff} = 1/2$ \cite{AbragamBleaney}. Indeed, based on INS studies, a single-ion model for \NCCF \ was suggested which identified the ground state as a $J_{\rm eff} = 1/2$ Kramers-doublet with strongly anisotropic $g$-factors \cite{Ross2017}. For comparison of this model with our ESR and magnetization measurements which probe the magnetic properties of the bulk, one has to take two further aspects of the crystal structure into account. First, the directions in which the F octahedra surrounding each Co$^{2+}$ ion are compressed differ for the four corners of the tetrahedra that form the magnetic sublattice of the pyrochlore system (see, for instance, Fig. 1 in Ref.~\cite{Ross2017}). Thus, there exist four distinct local-symmetry axes (denoted also as local $\langle111\rangle$ axes). Second, site disorder within the $A$ sublattice caused by the random distribution of Na and Ca \cite{Krizan2014} leads to random variations of the crystal fields acting on the Co ions and, consequently, to local variations in the $g$-tensor elements. Therefore, magnetization measurements essentially probe effective $g$-factors averaged over the different magnetic sites and projected onto the direction of the applied magnetic field, which might lead to a reduction of the experimentally observed anisotropy. Nevertheless, using the model proposed in Ref.~\cite{Ross2017} one might still anticipate effective $g$-factors larger than two. ESR as a local probe could in principle resolve individual Co ions with their respective $g$-tensors if they were not interacting. The $g$-factor of about 2 obtained for the dominant line L2 from frequency-dependent measurements at 30\,K evidences the strong exchange interactions between the Co spins in this system which average the local $g$-factor anisotropies in the ESR response.

Furthermore, if the resonance line observed in this temperature range were due to paramagnetic spins, the temperature dependence of the line's intensity would follow a Curie-law, i.e., it would show a smooth decrease with increasing temperature instead of the sudden disappearance observed. A possible explanation lies in different dynamical regimes of the spin system. In the paramagnetic state,  relaxation of spins could be fast due to the considerable spin-orbit coupling which leads to the observed $J_{\rm eff} = 1/2$ ground state. Such a fast relaxation would result in a very large linewidth which could, in addition to the small susceptibility \cite{Krizan2014}, prevent detection of the resonance line. Moreover, as it was stated in Ref.~\cite{Ross2016}, the first excited $J_{\rm eff} = 3/2$ states are thermally populated at room temperature which further enhances the spin relaxation in this temperature range. Upon cooling below temperatures comparable with $|\Theta_{CW}|$ of 140\,K spin fluctuations slow down as the coupling between spins increases, thereby leading to increasing relaxation times. Thus, it is reasonable to attribute the appearance of a resonance line at 40\,K to the formation of a spin state with reduced relaxation at the ESR time scale. We interpret the mode L2 associated with this phase as a gapless excitation of a cooperative paramagnetic state (referred to as ``$XY$ thermal spin liquid'' in Ref.~\cite{Ross2017}) present at these temperatures, which is, most likely, the precursor of the frozen state discussed in Ref.~\cite{Ross2016}. Consequently, the onset temperature of 40\,K, i.e., the temperature at which L2 starts to become visible in the spectra, characterizes the energy scale of the cooperative paramagnet. This is consistent with correlation lengths of the $XY$ correlations, i.e., correlations associated with the proposed short-range- ordered $XY$-type ground state \cite{Ross2016,Ross2017}, determined from an analysis of the magnetic pair distribution function (mPDF) up to 50\,K \cite{Frandsen2017}. They were shown to decrease down to $\sim$ 3\,\AA \ at 50\,K \cite{Frandsen2017} which is comparable with the nearest-neighbor Co-Co distances. Moreover, these fluctuations coexist with uncorrelated fluctuations above $\sim$~15\,K which is compatible with the scenario proposed here for the ESR mode L2 and a fast relaxation for temperatures above 40\,K in spite of the persistence of short-range correlations. Note that in Ref.~\cite{Ross2017} the temperature range for the existence of a cooperative paramagnetic phase is reported to extend even up to 140\,K, based on the absolute value of the Curie-Weiss temperature obtained for \NCCF. 

Another important finding of the ESR studies is the appearance of the mode L1 below 20\, K  and the related shift of spectral weight from L2 to L1 upon lowering the temperature. Interestingly, the observation of L1 coincides with the onset of changes in the spin-lattice and spin-spin relaxations times as seen in NMR measurements presented in Ref.~\cite{Sarkar2017}. The different character of the two excitation modes L1 and L2, and thus of the underlying correlated phases, reflects itself in the ESR measurements by the excitation gap found for L1 in contrast to the gapless behavior of L2. Although INS investigations revealed the existence of gapless low-energy excitations below $T_f \sim 2.4$\,K within their energy resolution of 0.17\,meV \cite{Ross2016} ($\sim$ 40\,GHz) the observed gap of L1 of about 59\,GHz could still be consistent with excitations of the increasingly frozen state, i.e., a cluster-spin-glass state. Moreover, we stress that the intensity of an ESR line is governed by the imaginary part of the dynamical spin susceptibility $\chi''(\omega, \vec{q} = 0)$ at zero wave vector $\vec{q}$. Consequently, ESR spectroscopy probes low-energy excitations at the $\Gamma$ point of the Brillouin zone, whereas the gapless excitations have been found in INS away from the zone center \cite{Ross2016}. Thus, although being potentially gapless at finite momentum transfers the excitations of the frozen state might still show a gapped behavior at the $\Gamma$ point.

In the temperature range within which ESR signals have been found, L2 did not vanish completely. Thus, one may conclude that between 3 and 20\,K there is a coexistence of two distinct magnetic phases: a cooperative paramagnetic phase with slowed down spin dynamics as compared to the paramagnetic state (leading to L2) and a spin-glass-like frozen phase which becomes more and more extended towards low temperatures (responsible for L1). 

\subsection{Magnetization}
\label{sec:discussion_Magnetization}
Despite an extreme $B$ over $T$ ratio in our magnetization experiment (58\,T, 2\,K), magnetic saturation cannot be achieved in consequence of the strong crystal electrical field (CEF) splitting in \NCCF. More surprisingly, measurements of the magnetization at pulsed magnetic fields along different crystal orientations did not show a pronounced magnetic anisotropy, as proposed by neutron-scattering experiments \cite{Ross2016,Ross2017}. In the following a possible explanation for these findings in terms of an empirical approach is discussed.

\begin{figure}[t]
	\centering
	\includegraphics[width=\columnwidth]{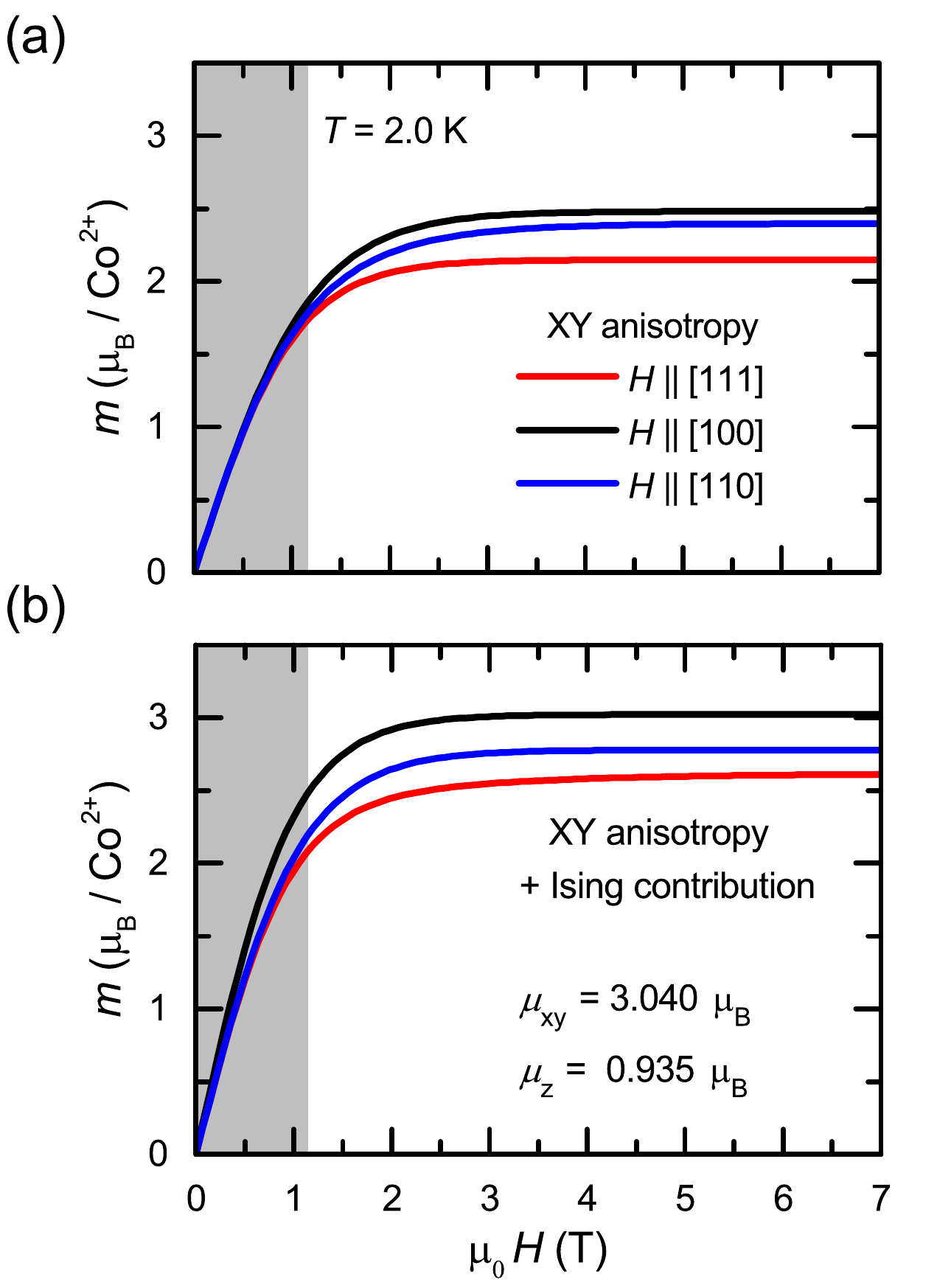}
	\caption{(a) Calculated magnetization of non-interacting spins with pure  $XY$ anisotropy on a lattice of corner-sharing tetrahedra. (b)  Behavior for non-interacting spins with an additional small Ising component. Gray shaded areas mark the field range which corresponds to the range of 0 - 55\,T if interactions between the spins are taken into account empirically.}
	\label{fig:NCCF_04}
\end{figure}

The magnetization $m$ of non-interacting $XY$ spins on a lattice of corner-sharing tetrahedra can be calculated by the projection of each spin onto the field direction
\begin{equation}
m (\vec{\mu}) = \dfrac{1}{4} \sum_{i=1}^{4} \mu_\mathrm{xy} |\vec{n}_i \times \vec{h}| \tanh \left( \frac{\mu_\mathrm{xy} |\vec{n}_i \times \vec{h}| B}{\kB T} \right) \, , 
\label{eq:Magnetisierung_Tetraeder_XY}
\end{equation}
where $\vec{n}_i$ are the normalized vectors of the four local \eee -directions, $\vec{h} = \vec{H}/|\vec{H}|$ is the direction of the field and $B = \mu_0 |\vec{H}|$.
The cross product $|\vec{n}_i \times \vec{h}|$ gives the projection of the local  $XY$ plane onto the field direction.
The resulting curves are shown in Fig. \ref{fig:NCCF_04}(a) for a moment of $\mu_\mathrm{xy}$ = 3.04 \muB at $T$ = 2~K. The particular value of $\mu_\mathrm{xy}$ is chosen according to Ref.~\cite{Ross2017}.

Additionally to the leading moment within the local $XY$ plane, a small Ising contribution of $\mu_\mathrm{Ising}$~= 0.935~\muB was found by neutron scattering  \cite{Ross2017}. This leads to an additional component of the magnetic moment for each magnetic site,

\begin{equation}
\sum_{i,j = \pm} \langle i|\hat{\vec{L}}+2\hat{\vec{S}}|j \rangle\vec{H} =  \pm \left( \mu_\mathrm{xy}H_\perp + \mu_\mathrm{z}H_\| \right) \, ,
\end{equation}
following the definition of the moments $\mu_\mathrm{xy}$ and $\mu_\mathrm{z}$ given by Ross \etal \cite{Ross2017}. Consequently, Eq.~(\ref{eq:Magnetisierung_Tetraeder_XY}) can be extended by not only considering the projection onto the $XY$ plane, $H_\perp = (\vec{n}_i \times \vec{h})|\vec{H}|$, but also taking into account the projection of the Ising contribution onto the field, $H_\| = (\vec{n}_i \cdot \vec{h})|\vec{H}|$ which leads to
\begin{multline}
	\label{eq:Magnetisierung_Tetraeder_XY+Ising}
	m (\vec{\mu}) = \dfrac{1}{4} \sum_{i=1}^{4} \left( |\mu_\mathrm{xy} \vec{n}_i \times \vec{h} + \mu_\mathrm{Ising} \vec{n}_i \cdot \vec{h}| \right) \\
	\times \tanh \left( \frac{|\mu_\mathrm{xy} \vec{n}_i \times \vec{h} + \mu_\mathrm{Ising} \vec{n}_i \cdot \vec{h}| B}{\kB T} \right)
	\, .
\end{multline}
The expected magnetization is, thus, altered [Fig. \ref{fig:NCCF_04}(b)] but saturation is still expected at small field whereas for high fields different values are expected for the saturation magnetization. Note, that higher CEF levels are neglected in this consideration because the first excited state is split by an energy $\Delta_\text{CEF}$ corresponding to $\mu_0 H = \Delta_\text{CEF} / \mueff = 150$~T \cite{Ross2017}.

To now consider the interaction of the spins, the temperature is replaced by an effective temperature \mbox{$T \rightarrow T_\mathrm{eff} = T + T_0$}.
In this phenomenological approach $k_\mathrm{B}T_0$ is the energy scale of the average local frustrated interaction.
This approach is similar to the use of a modified Brillouin function in the work of Heiman \etal \cite{Heiman1984} and Cisowski \etal \cite{Cisowski1999}. 
In Fig.~\ref{fig:NCCF_02}, this function is fit to the experimental results. For clarity, fit results are shown separately in the inset in Fig.~\ref{fig:NCCF_02}(b) for the case of measurements with different orientations as they show an isotropic behavior and overlap with the experimental data. Note, that $T_0$ is the only free parameter of this fit. For $\mu_\mathrm{xy}$ and $\mu_\mathrm{Ising}$ the discussed values from Ross \etal \cite{Ross2017} were used. Because $T_0$ is assumed to be field independent, we consider the fact that the frustration is established also in zero field.

\begin{table}[t]
	\centering
	\caption{Strength of the phenomenological parameter $T_0$ for the different measurements shown in Fig. \ref{fig:NCCF_02}.}
	\setlength\extrarowheight{0.5pt}
	\begin{ruledtabular}
		\begin{tabular}{c c c}
			Orientation & $T$ (K) & $T_0$ (K)\\
			\hline
			$H\|$[100] & 2 & 140\\
			$H\|$[111] & 2 & 111\\
			$H\|$[110] & 2 & 118\\
			\hline
			$H\|$[110] & 35 & 108\\
			$H\|$[110] & 85 & 110\\
			$H\|$[110] & 129 & 115\\
		\end{tabular}
	\end{ruledtabular}
	\label{tab:NCCF_T0s_aus_Fit}
\end{table}

The different values obtained for $T_0$ are summarized in Table~\ref{tab:NCCF_T0s_aus_Fit}. The magnetization for different temperatures in the orientation $H\|$[110] can be described by a nearly temperature-independent value of the internal interactions of $T^\mathrm{[110]}_0 = 113 \pm 2$~K. Whereas the magnetization for the orientation $H\|$[111] can be described using a similar value, for $H\|$[100] a slightly higher parameter of $T^\mathrm{[100]}_0$~= 140~K is necessary. These $T_0$ values are in the range of the absolute value of the Curie-Weiss temperature. By that, $T_0$ indeed proves to be a representative quantity for the frustrated local spin-spin interactions and, to a limited extent according to its simplified definition, for their anisotropy, too. Caused by the use of an effective temperature, the magnetization at 2\,K in the range of 0 - 55\,T corresponds to a field range of only 0 - 1.15\,T for a non-interacting system  (gray shaded areas in Fig.~\ref{fig:NCCF_04}). In this region, the expected magnetization is rather similar for all three directions which could lead to the impression of isotropic spins.

It should be mentioned that this approach is phenomenological and all local variations of the CEF as well as the exchange and dipole interactions are treated by an average and constant energy scale $T_0$. This approach can describe how a local $XY$ anisotropy could be masked within magnetization measurements as a result of strong magnetic frustration. Note, that the values of the local moments used for determination of $T_0$ are based on the highly anisotropic ionic $g$-tensors reported in Ref.~\cite{Ross2017}. While they still allow to explain the isotropic behavior found in field-dependent magnetization measurements, they are in contrast to the effective $g$-factors close to 2 determined from ESR experiments which, in turn, could yield an isotropic magnetization as well. The effective $g$-factors result, most likely, from an averaging over different local magnetic sites induced by exchange interactions, see the discussion above. Therefore, we suggest that both experimental signatures for isotropic magnetic behavior obtained in this work from field-dependent magnetization and ESR measurements are caused by strong magnetically frustrated interactions despite the potential presence of a local anisotropy of individual Co ions.

\section{Conclusions}
\label{sec:conclusions}

In conclusion, we presented a combined high-field ESR and magnetization study of the magnetic properties of the pyrochlore compound \NCCF. ESR experiments in the temperature range 3 - 40\,K, i.e., above the freezing temperature $T_f$, revealed a low-energy excitation spectrum whose components could be assigned to two distinct coexisting magnetic phases. Below 40\,K, the ESR response is dominated by a gapless mode which we ascribe to a cooperative paramagnetic phase. By further cooling, spectral weight is progressively shifted towards a gapped mode associated with the spin-glass-like phase which is expected to extend over the whole crystal below $T_f$. For these two modes, effective $g$-factors close to 2 were determined. This finding is consistent with the isotropic field-dependent magnetization up to 58\,T. In addition to the observed isotropy, these measurements did not show signatures of saturation even at 2\,K. An empirical model was proposed explaining the isotropic behavior in the magnetization experiments by taking into account anisotropic ionic $g$-tensors reported in literature \cite{Ross2017} and introducing an effective temperature $T_0$. However, this approach could not provide a direct link between the anisotropic $XY$-like $g$-tensors of the single-ion model discussed in Ref.~\cite{Ross2017} and the apparently isotropic effective $g$-factors obtained in the present ESR experiments. The latter is most likely related to the strong geometrical frustration present in the title compound. Thus, the presented work should motivate further theoretical studies investigating details of the magnetic ground state of \NCCF and its low-energy excitations, which we showed to be more complex than expected based on a pure single-ion model.

\section*{Acknowledgments}
We would like to thank C.G.F. Blum for technical support. This research is supported by the Deutsche Forschungsgemeinschaft (DFG) through SFB 1143 and KA1694/8-1. We also acknowledge the support of HLD at HZDR, member of the European Magnetic Field Laboratory (EMFL). The crystal-growth work at Princeton University was supported by the US DOE Division of Basic Energy Sciences, grant DE-FG02-08ER46544.

\appendix

\section{Samples used in this study}
\label{sec:sample_details}

\begin{figure}[h]
	\includegraphics[width=\columnwidth]{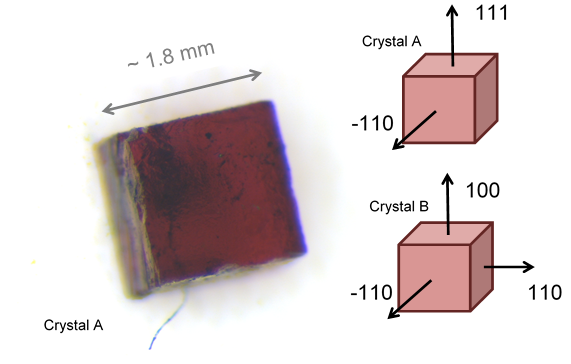}
	\caption{One of the samples used for high-field magnetization measurements (left). Symmetry axes along which the measurements have been carried out are schematically shown for the two cube shaped samples on the right side.}
	\label{fig:picture_sample}
\end{figure}

For the investigations presented in this work different samples cut from one large single crystal were used. The latter was reported and characterized previously in Ref.~\cite{Krizan2014}. Magnetization measurements were conducted on two crystals which were cut to cube-shaped samples with lengths of a side of about 1.8 mm in order to reduce the difference in demagnetization effects as much as possible and allow a comparison between magnetization measurements along different crystallographic directions. One of these samples is shown on the left side of Fig.~\ref{fig:picture_sample}. On the right side, schematic pictures of both samples including information about their symmetry axes are given.

\begin{figure}[t!]
	\includegraphics[width=\columnwidth]{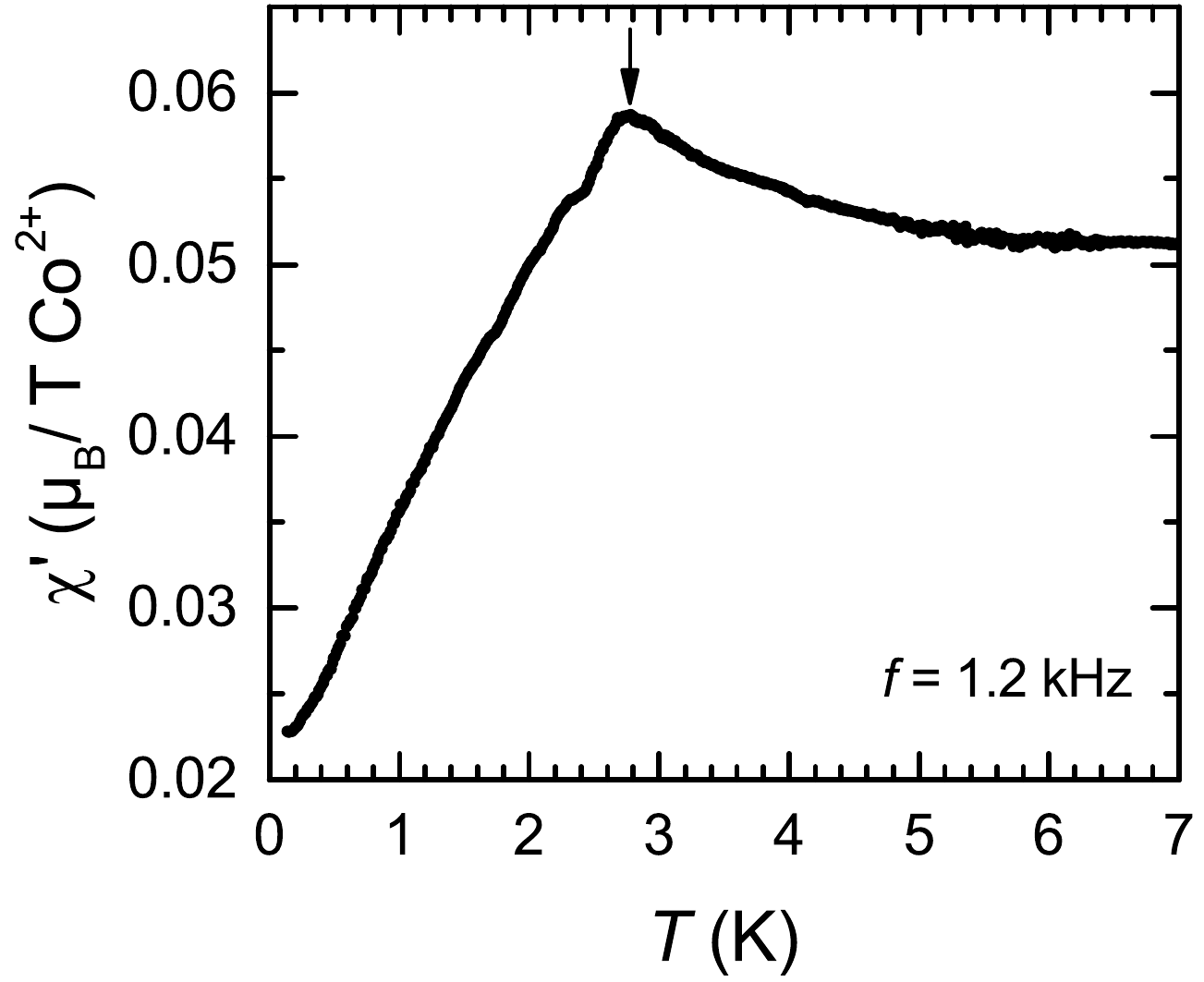}
	\caption{Low-temperature measurement of the real part $\chi'$ of the dynamical susceptibility measured at a frequency \mbox{of 1.2\,kHz} without applying an external magnetic field. The arrow indicates the maximum in $\chi'(T)$ which evidences the transition into the frozen state.}
	\label{fig:dyn_susc}
\end{figure}

Before performing the high-field experiments the unchanged quality of the samples, as compared to the initial studies in Ref.~\cite{Krizan2014}, was checked by measurements of the static magnetization in low  fields. The results of these experiments are in agreement with those given in Ref.~\cite{Krizan2014} and are not shown here. In addition, the temperature dependence of the dynamical susceptibility was measured in the low-temperature region down to 0.15\,K with a frequency of 1.2\,kHz.
As it is shown in Fig.~\ref{fig:dyn_susc}, the real part $\chi'(T)$ of the dynamical susceptibility features a maximum indicating the freezing temperature which is shifted to a slightly higher temperature of 2.77\,K as compared to the static measurements, in line with the frequency dependence of this maximum reported in Ref.~\cite{Krizan2014}. Below the freezing temperature no evidence for further transitions, for instance to a state with long-range order, is found down to the lowest temperature of 0.15\,K.

For ESR experiments, a larger crystal was used as for the cubic samples the sample mass was too low to observe all ESR lines with sufficient intensity (note that the intensity of a resonance line is proportional to the number of spins contained in the sample).

\clearpage
\bibliography{literature_NaCaCo2F7}

\end{document}